\documentclass{revtex4}
\usepackage{amsmath}

\usepackage{epsfig,epsf}
\usepackage{graphicx}
\usepackage{grffile} 
\usepackage{pdfpages}

\usepackage{amsmath,amsfonts,amssymb,amsthm,nccmath,latexsym,mathtools} \usepackage[mathscr]{euscript}
\usepackage[hidelinks]{hyperref}
\hypersetup{
    colorlinks,
    citecolor=black,
    filecolor=black,
    linkcolor=black,
    urlcolor=black
}
\hypersetup{linktocpage}

\newcommand{\be}{\begin{equation}}
\newcommand{\ee}{\end{equation}}

\usepackage{amstext} % for \text macro
\usepackage{array}   % for \newcolumntype macro
\newcolumntype{L}{>{$}l<{$}} % math-mode version of "l" column type
\usepackage{epsfig,epsf}
\usepackage{graphicx}
\usepackage{grffile} 
\usepackage{braket}
\usepackage{slashed}

\newcommand{\LL}{\mathcal{L}}

\newcommand{\del}{\partial}

\usepackage{pdfpages}

\usepackage{amsmath,amsfonts,amssymb,amsthm,nccmath,latexsym,mathtools} \usepackage[mathscr]{euscript}
\usepackage[hidelinks]{hyperref}
\hypersetup{
    colorlinks,
    citecolor=black,
    filecolor=black,
    linkcolor=black,
    urlcolor=black
}

\begin{document}

%\hoffset-1cm
 
% Yale printer values

%\draft{BI-TP 2005/19}
%\preprint{BI-TP 2005/19, CERN-PH-TH-2005/99}
\title{2+1 d Georgi Glashow Model Near Critical Temperature.}
\author{Candost Akkaya$^1$, Ibrahim Burak Ilhan$^{1}$ and Alex Kovner$^{1}$}

\affiliation{
$^1$ Physics Department, University of Connecticut, 2152 Hillside
Road, Storrs, CT 06269-3046, USA}
\date{\today}

\begin{abstract}
We study correlation functions of magnetic vortices $V$ and Polyakov loop $P$ operators in the 2+1 dimensional Georgi-Glashow model in the vicinity of the deconfining phase transition. In this regime the (dimensionally reduced)   model is mapped onto a free theory of two massive Majorana fermions. We utilize this fermionic representation to explicitly calculate the expectation values of $V$ and $P$ as well as their correlators. In particular we show that the $VV$ correlator is large, and thus the anomalous breaking of the magnetic $U(1)$ symmetry is order one effect in the near critical region. We also calculate the contribution of magnetic vortices to the entropy and the free energy of the system.

\end{abstract}
\maketitle
%%%%%%%%%%%%%%%%%%%%%%%%%%%%%%%%%%%%%%%%%%%%%%%%%%%%%%%%%%
\section{Introduction}
Understanding of the deconfining phase transition in confining gauge theories is one of the outstanding problems of modern field theory. In a certain sense it is even more interesting to try and understand not the transition itself, but the near critical region. While there exist universality arguments that pertain to critical behavior itself \cite{svet}, they do not strictly speaking apply at temperatures away from $T_c$. On the other hand we know from lattice studies that nonperturbative strongly interacting physics is relevant at temperatures significantly higher than $T_c$. 

 In QCD for example it has been known for a long time that the free energy and entropy significantly deviate from their expected high $T$ behavior up to relatively high temperatures \cite{karsch}. 
It has been conjectured that this discrepancy in large measure is due to abundance in the thermal ensemble of topological excitations (magnetic vortices or magnetic monopoles) which are believed to be responsible for confinement at $T=0$. These topological objects affect thermodynamic properties as well as correlation functions, and their effect dies away only at $T\gg T_c$.

Needless to say that solving a near critical QCD is a very hard problem.  It is thus worthwhile to turn to simpler theories that share some salient features with QCD to try and get some understanding, or at least an inspiration. One such ``fall back" model is the Polyakov's confining Georgi-Glashow model in 2+1 dimensions \cite{polyakov}. It is confining at zero temperature with string tension and mass gap calculable in weak coupling regime. Moreover, it undergoes a deconfining phase transition at finite temperature. A number of years ago it has been understood how to quantitatively study this transition \cite{dkkt}. The deconfining phase transition in this model is understood simply as the symmetry breaking transition of the magnetic $Z_2$ symmetry.  The magnetic vortex operator, or 't Hooft loop is the local order parameter for this symmetry \cite{kr}.  It was  shown directly in \cite{dkkt} that the critical theory is a two dimensional Ising model consistent with the universality arguments.

 The critical Ising model is equivalent to the theory of a single Majorana fermion. In the framework of the Georgi-Glashow model 
the Majorana theory was derived directly from the original partition function in \cite{dkkt}. Moreover, even away from criticality the dimensionally reduced partition function is exactly transformed into a two dimensional fermionic model. Close to the critical temperature in the interesting near critical region, the fermionic model is noninteracting. It thus becomes possible in this confining theory to perform controlled calculations close to critical temperature. 

Although the fact that the near critical Georgi Glashow model is equivalent to a noninteracting theory has been recognized in \cite{dkkt}, this equivalence has not been utilized for quantitative study. The aim of the present paper is to demonstrate how certain correlation functions and thermodynamic quantities in the Georgi-Glashow model can be calculated using the aforementioned fermionic representation. In particular we  calculate the expectation value of the Polyakov loop and magnetic vortex operator ('t Hooft loop) as well as their  correlation functions  in the near critical regime. We also calculate the free energy and the entropy in the same regime.

 The plan of this paper is the following. In Sec.2 we recap the relation of the near critical Georgi-Glashow model with two dimensional theory of noninteracting fermions. In Sec. 3 we use the fermionic representation to calculate the expectation value of the two order parameters of the GG model - the magnetic vortex operator ('t Hooft loop) and Polyakov loop. We show that the former has vanishing expectation value above $T_c$ due to appearance of a normalizable zero mode of the fermionic Dirac operator. The latter suffers similar fate at $T<T_c$. This result is of course not new. Rather the purpose of this exercise is to demonstrate how the known results are obtained using fermionized formulation.
In Sec. 4  we calculate  the vortex-antivortex   correlation function above the critical temperature. In Sec. 5 we consider the vortex-vortex correlator. We demonstrate that close to the phase transition this correlation function is large, so that the anomalous breaking of the magnetic $U(1)$ symmetry is an order one effect. We also show that within our approximation there is a degeneracy between the scalar and pseudoscalar correlation length, even though the numerical value  the pseudoscalar correlator is much smaller. In Sec. 6 we calculate the free energy end entropy of the system. We pay particular attention to the question how the entropy and free energy are modified if one neglects the effect of vortices, or more accurately of the anomalous magnetic $U(1)$ symmetry breaking in the thermal ensemble. Finally we conclude with a short discussion  in Sec.7. 

\section{The theory close to the deconfining phase transition.}
\subsection{The model and the relevant degrees of freedom.}

The 2+1 dimensional Georgi-Glashow model is the $SU(2)$ gauge theory of an adjoint Higgs field:
\begin{equation}
L=-\frac{1}{4}F^a_{\mu\nu}F^{a\mu\nu}+\frac{1}{2}(DH)^2-\lambda(H^2-v^2)^2
\end{equation}
The Higgs expectation value is assumed to be much larger than the gauge coupling constant (implicit in the covariant derivative) $v^2\gg g^2$.

Perturbatively the spectrum of the theory contains a massless ``photon" as well as massive charged gauge bosons $W_\pm$ with masses of order $M^2_W=g^2v^2\gg g^4$. Additionally there is a neutral massive Higgs boson. This last particle  will play no role in our discussion  albeit we will assume that it is light enough (see later).

It is well known that the nonperturbative effects lead to nonperturbative mass generation of the photon and confinement of the charged $W$ bosons with an exponentially small string tension. The effective low energy description is given by the Polyakov effective action which involves one scalar ``dual photon" field $\chi$ \cite{polyakov}
\be\label{l3}
L=\frac{g^2}{8 \pi^2 }(\partial_\mu \chi)^2+ \tilde\zeta \cos{2 \chi}
\ee
with the ``monopole fugacity"  $\tilde \zeta\propto e^{-4\pi M_W/g^2}$.
The mass of the dual photon is obviously $m^2_{ph}=\frac{4\pi^2\tilde \zeta}{g^2}$. The string tension can be calculated classically and is parametrically $\sigma\sim g^2m_{ph}$.

This Lagrangian eq. (\ref{l3}) is in fact the direct analog of effective chiral Lagrangian of QCD, as it can be written in terms of the local order parameter field, the magnetic vortex field \cite{kr}
\be V(x)=\left(\frac{g^2}{8\pi^2}\right)^{1/2}e^{-i\chi}\label{vo}\ee
\be L=|\partial_\mu V|^2+m^2_{ph}(V^2+V^{*2})\ee
The nonvanishing expectation value of $V$ signifies spontaneous breaking of the magnetic $Z_2$ symmetry $V\rightarrow -V$. In the absence of the vortex mass term the magnetic symmetry is enhanced to the full $U(1)$ rotation group. The monopole induced mass term is small and is in many respects analogous to the instanton induced anomaly of the $U_A(1)$ axial symmetry in QCD \cite{kr}.

As with any theory with spontaneous breaking of a global symmetry, one expects symmetry restoration when the temperature is high enough. Indeed as shown in \cite{dkkt} at $T_c=\frac{g^2}{4\pi^2}$ the Georgi-Glashow model undergoes a $Z_2$ restoring phase transition\footnote{The exact vale of the critical temperature depends on the Higgs mass \cite{sonkovchegov}. The value quoted in the text is the correct one in the regime when the Higgs particle is lighter than the vector bosons, but much heavier than the scale set by the gauge coupling constant $g^2\ll M_H\ll M_W$. In this paper for simplicity we will only consider this regime.}. This phase transition is the Georgi-Glashow incarnation of  the deconfining phase transition, as it is accompanied with the appearance of a nonvanishing expectation value of the Polyakov loop \cite{dkkt}.

Since the value of the critical temperature is much greater than the mass scale at $T=0$,  dimensional reduction is valid in this model starting at temperatures much lower than $T_c$. As was shown in \cite{dkkt} the proper (Euclidean) dimensionally reduced theory close to criticality is given by the Lagrangian

\be \mathcal{L}=\frac{g^2}{8 \pi^2 T}(\partial_\mu \chi)^2+ \zeta \cos{2 \chi}+\mu \cos{\tilde{\chi}},\ee

 Here $\zeta= \tilde\zeta/T$ while  $\mu$ is proportional to the fugacity of the heavy charged vector bosons $\mu\propto T^2e^{-\frac{M_W}{T}}$. 
 The dimensionally reduced theory is valid on distance scales greater than the inverse temperature, and is therefore defined with the UV cutoff $\Lambda\approx T$.
 
 The dual field $\tilde{\chi}$ is defined as:

\be\label{chi}
 i\partial_\mu \tilde{\chi} = \frac{g^2}{2\pi T}\epsilon_{\mu\nu}\partial^\nu \chi; \ \ \ \ \ \ \  i\tilde{\chi}(x) = \frac{g^2}{2\pi T}\int^x_C dx_\mu\epsilon_{\mu\nu}\partial_\nu \chi.\ee
where the contour of the integration $C$ starts at an arbitrary point at infinity and ends at the point $x$.  

The dual field $\tilde \chi$ is related to the (fundamental) Polyakov loop operator of the Georgi-Glashow model:
\be tr\ {\cal P}=\cos\frac{\tilde\chi}{2}\ee
In the following with a slight abuse of notation we will use the abelianized Polyakov loop operator, which is more appropriate to the  weak coupling regime
\be P=e^{-i\frac{\tilde \chi}{2}}\ee
The relation eq.(\ref{chi}) strictly speaking is valid to leading order in $\mu$ and $\zeta$, since the curl of the left hand side and the divergence of the right hand side vanish. The more general duality relation valid for finite $\mu$ and $\zeta$ is
\be\label{dp} iP^*\partial_\mu P =-\frac{4\pi^2T}{g^2}\epsilon_{\mu\nu}V^*\partial^\nu V
\ee
The presence of the factor $i$ on the left had side of eqs.(\ref{chi}) and (\ref{dp}) is consistent with the fact that in Euclidean space the zeroth component of vector potential is imaginary. 

In the following we will fermionize this theory, which is most straightforwardly done in Minkowski space. There the analogous relation looks more natural and reads
\be\label{dpm} P^*\partial_\mu P =\frac{4\pi^2T}{g^2}\epsilon_{\mu\nu}V^*\partial^\nu V
\ee

\subsection{Fermionization.}
The dimensionally reduced theory can be conveniently rewritten in fermionic form. To do this we use the standard bosonization techniques of Coleman and Mandelstam \cite{boson1, boson2}.
Let us define for convenience a scaled field
\be \chi = \phi \sqrt{\frac{4\pi^2 T}{g^2}} \equiv \phi \frac{\beta}{2} \label{beta}.\ee 

\be  i\tilde{\chi} = \frac{4\pi}{\beta} \int dx_\mu \epsilon_{\mu\nu}\partial_\nu \phi \label{vor}.\ee

To perform fermionization {\it a la} Mandelstam we make the rotation to Minkowski space
\be x \rightarrow i t.\ee
For eq. (\ref{vor}) in Minkowski space, we have:
\be i\tilde{\chi} = \frac{ 4 \pi i}{\beta} \int^y d\lambda \dot{\phi}(\lambda).\ee
 We define the spinors in the standard way \cite{boson2}:
\begin{align}\begin{split}
\psi_1(y) &=A \exp\left({\frac{-2\pi i}{\beta}\int_{-\infty}^y d \lambda \dot{\phi}(\lambda) - \frac{1}{2} i \beta \phi(y)}\right) = Ae^{-i(\frac{1}{2} \tilde{\chi} + \frac{\beta}{2}\phi)}%= P V^\dagger
,  \\
\psi_2(y) &=-i A \exp\left({\frac{-2\pi i}{\beta}\int_{-\infty}^y d \lambda \dot{\phi}(\lambda) + \frac{1}{2} i \beta \phi(y)}\right)=  -i Ae^{-i( \frac{1}{2}\tilde{\chi} - \frac{\beta}{2}\phi)}%=-iPV
. \label{psi11}
\end{split}\end{align}
where $A$ is a normalization constant.
With our choice of phases in eq.(\ref{psi11}) the representation of Dirac matrices in the fermionized theory is
\be \gamma^0=\sigma^1, \mbox{ }\gamma^1 = i\sigma_2, \mbox{ }\gamma^5=\gamma^0\gamma^1=-\sigma_3.\ee

 The currents in the two representations are related as \cite{boson2}\footnote {The seemingly noncovariant form of this relation is to do with our use of Hamiltonian formalism and therefore a noncovariant way of UV regularization like in \cite{boson2}. In a covariant Lagrangian formalism the fermionic current is given by the usual $\bar \psi\gamma^\mu\psi$.} : 
\be \label{current}j^\mu = -\frac{\beta}{2\pi}\epsilon^{\mu\nu}\partial_\nu \phi=\left(\delta^\mu_0+\frac{\beta^2}{4\pi}\delta^\mu_1\right)\bar{\psi}\gamma^\mu \psi,\ee

 Note that in terms of the vortex and Polyakov operators

\be \psi_1\propto  P V , \ee
\be \psi_2\propto  -iP V^\dagger. \ee

It is convenient to express $V$ and $P$ directly in terms of the fermionic degrees of freedom. Using eq.(\ref{current}) we obtain

%The relevant current for $V(x)$ is

%\be j^0 = -\frac{\beta}{2\pi}\phi'=\bar{\psi}\gamma^0 \psi,   \mbox{		}   [j^0(x),\psi(y)]=-\psi(x)\delta(x-y),\ee

%which gives the phase of the operator:

\be V(x) =\sqrt{\frac{8\pi}{g^2}}e^{-i\frac{\beta}{2}\phi(x)}=\sqrt{\frac{8\pi}{g^2}}\exp{\left(-i\frac{\beta}{2}\int^x_{-\infty}d\lambda \phi'(\lambda)\right)}=\sqrt{\frac{8\pi}{g^2}} \exp{\left(i\pi\int^x_{-\infty}\bar{\psi}\gamma^0\psi\right)},\ee.

%We calculate the current for $P(x)$ in a similar way:

%\be j^1 = \frac{\beta}{2\pi}\dot{\phi}=\frac{\beta^2}{4\pi}\bar{\psi}\gamma^1\psi,   \mbox{		}   [j^1(x),\psi(y)]=-\frac{\beta^2}{4\pi}\sigma_3\psi(x)\delta(x-y),\ee

%and

\be P(x) =e^{-i\frac{1}{2}\tilde{\chi}(x)}=\exp{\left(-i\frac{2\pi}{\beta}\int^x_{-\infty}d\lambda \dot{\phi}(\lambda)\right)}= \exp{\left(-i\pi\int^x_{-\infty}\bar{\psi}\gamma^1\psi\right)}.\ee

We have the explicit forms of the vortex and Polyakov loop operators in terms of the spinors as (we rescale the vortex operator for convenience):

\be V(x) = \label{vor1}  e^{i\pi \int^x_{-\infty}dx \left( \psi^\dagger_1\psi_1 + \psi^\dagger_2\psi_2 \right) },\ee
\be P(x) =  \label{pol} e^{i\pi \int^x_{-\infty}dx \left(\psi^\dagger_1\psi_1 - \psi^\dagger_2\psi_2\right)}.\ee

The usual bosonization rules for the fermonic bilinears then give the following fermionized form of the dimensionally reduced model:

\be  \mathcal{L}= \bar{\psi}i\gamma_\mu \partial^\mu \psi+ \frac{1}{2}\lambda j^\mu j_\mu+m_1\bar{\psi}\psi + m_2\left(\psi_2\psi_1 + \psi^\dagger_1\psi^{\dagger}_2 \right).\ee
where
\be \frac{\lambda}{\pi}=\frac{4\pi}{\beta^2}-1; \ \ \ \ \ \  m_1\propto \zeta/T; \ \ \ \ \ \ m_2\propto\mu/T\ee
Note that at criticality ($\beta^2=4\pi$) the coupling $\lambda$ vanishes, and the theory describes two free Majorana fermions. Additionally at critical temperature $m_1=m_2$ so that one of the  fermions is massless, while the other one is massive. Close to criticality the most important relevant perturbation is $m_1-m_2$. This means that when calculating various correlation functions close to critical temperature we can still treat the two fermions as noninteracting.  In fact, since the perturbation theory in $\lambda$ is infrared finite, we can neglect the effects of the fermionic coupling in the whole region $|T-T_c|/T\ll 1$, and consider the fermions to be free. In the rest of this paper we will set the four fermi coupling to zero $\lambda=0$.

\subsection{Back to Euclidean space.}
The last step in the fermionization transformation is to transform the path integral back into Euclidean space, where it was originally formulated. To do this we use the Wick rotation
 \be t \rightarrow -ix \label{wick}.\ee 
 This rotation affects the spinors  in the following way:

\begin{align}\begin{split}
\psi^E_1(y) &= A \exp\left({\frac{2\pi }{\beta}\int_{-\infty}^y d \lambda \phi'(\lambda) - \frac{1}{2} i \beta \phi(y)}\right)  \equiv \xi_1,\\
(\psi^\dagger_1(y))^E&= A \exp\left({\frac{-2\pi }{\beta}\int_{-\infty}^y d \lambda \phi'(\lambda) + \frac{1}{2} i \beta \phi(y)}\right) \equiv \xi^\dagger_2,\\
\psi^E_2(y) &=-i A \exp\left({\frac{2\pi }{\beta}\int_{-\infty}^y d \lambda \phi'(\lambda) + \frac{1}{2} i \beta \phi(y)}\right) = -i\xi^\dagger_1,\\
(\psi^\dagger_2(y))^E&=i A \exp\left({\frac{-2\pi }{\beta}\int_{-\infty}^y d \lambda \phi'(\lambda) - \frac{1}{2} i \beta \phi(y)}\right)  = i \xi_2,\\
\end{split}\end{align}
 The Lagrangian in  Euclidean space becomes:

\be \mathcal{L}_E=   -\left(\xi^\dagger_2\left(\partial_x -i  \partial_y\right)\xi_1 + \xi^\dagger_1\left(\partial_x +i  \partial_y\right) \xi_2\right)
-i m_1\left( \xi^\dagger_2\xi^\dagger_1 - \xi_2\xi_1 \right)-i m_2\left( \xi^\dagger_1\xi_1-\xi^\dagger_2\xi_2\right) \label{elag},\ee

while the Euclidean fermionized form of the vortex and Polyakov operators is
\be V(y)= \exp\left({i\pi\int^y(\xi^\dagger_2\xi_1 -\xi^\dagger_1\xi_2)}\right)\label{ve},\ee
\be  P(y)= \exp\left({i\pi\int^y(\xi^\dagger_2\xi_1 +\xi^\dagger_1\xi_2)}\right)\label{epol}.\ee
%Note, that consistently with our originl definition, $V$ is unitary while $P$ is Hermitian.

The line integral in the definition of the operators can be written in the following suggestive form  :

\be \int^y_{-\infty}  d\lambda f(\lambda) = \int d^2z \delta(z_1)\Theta(y-z_2)f(z) \label{sing}\ee

Defining the polar angle  as:
\be \theta = \arctan{\frac{z_1}{-z_2}} \label{angledef}\ee
%
%\begin{figure}[b]
%
%\begin{tikzpicture}[thick]
%\coordinate (O) at (0,0);
%\coordinate (K) at (3,0);
%\coordinate (L) at (0,3);
%\coordinate (M) at (-3,0);
%\coordinate (N) at (0,-3);
%\coordinate (r) at (2,-2);
%\draw (O)--(K);
%\draw (O)--(L);
%\draw (O)--(M);
%\draw (O)--(N);
%\draw (O)--(r);
%\draw [dashed]  (O)--(K) node[pos=0.5,xshift=1.7cm] { $z_1$};
%\draw [dashed]  (O)--(L) node[pos=0.5,yshift=1.7cm] { $z_2$};
%
%
%
%\pic["$\theta$", draw, ->, angle eccentricity=1.2, angle radius=1cm]
%    {angle=N--O--r};
%
%
%\end{tikzpicture}
%\caption{The angle $\theta$ defined.} \label{anglefig}
%\end{figure}
\begin{figure}[h]
	\caption{The angle $\theta$ defined.}\label{anglefig}
	\includegraphics[scale=0.8]{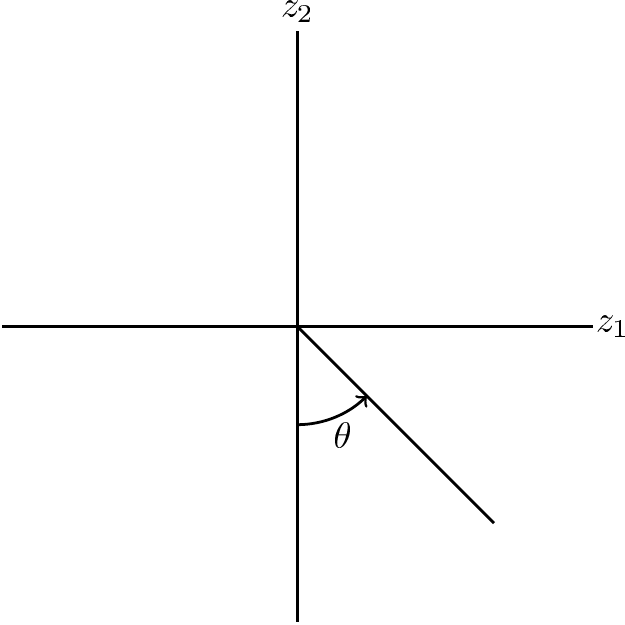}
\end{figure}
we have:

\be \partial_\mu \theta = \frac{\epsilon_{\nu\mu} z_\nu}{z^2}- 2\pi n_\mu \delta(z_1)\Theta(y-z_2) \equiv A_\mu- 2\pi n_\mu \delta(z_1)\Theta(y-z_2)  \label{tur}\ee

In our convention, the angle jumps by $2\pi$ across the negative $y$ axis. We summarize some useful formulae consistent with our definition of $\theta$ in the appendix.
The integral in eq.(\ref{sing}) as well as in eqs.(\ref{ve}) and (\ref{epol}) runs along the cut of the polar angle function. The position of this cut is arbitrary and should not affect any physical observable. Indeed as proved in \cite{dkkt}, the correlators of $V$ and $P$  do not depend on its position. This independence on the position of the cut will serve as a consistency check on our results.

\subsection{On $\langle P\rangle$ and $\langle V\rangle$.}

Our next order of business will be to calculate the expectation values of $V$ and $P$ as a function of temperature near criticality. Here we have to deal with the following problem. The fermionized Euclidean Lagrangian is real $\mathcal{ L^*}=\mathcal{ L}$, and so is the Polyakov operator. In fact the expectation value of $P$ is given by a Gaussian fermionic path integral
\be  <P(x)> =  \int D \xi D\bar{\xi} e^{-\int\mathcal{L}_P}.\label{action}\ee
with
\be \mathcal{L}_P=   -\left(\xi^\dagger_2\left[\partial_x -i  \partial_y -i\pi \delta(x)\Theta(-y)\right]\xi_1 + \xi^\dagger_1\left[\partial_x +i  \partial_y - i\pi\ \delta(x)\Theta(-y)\right] \xi_2\right)
-i m_1\left( \xi^\dagger_2\xi^\dagger_1 - \xi_2\xi_1 \right)-i m_2\left( \xi^\dagger_1\xi_1-\xi^\dagger_2\xi_2\right). \label{actionv}\ee
Since $\mathcal{L_P}$ is real, the expectation value of $P$ is thus simply given by the determinant of the modified Dirac operator.

The situation is somewhat different with the vortex operator. The operator eq.(\ref{ve}) is unitary. Thus even though formally its  expectation value is also given by a Gaussian path integral, the corresponding ``action" is complex. Thus one cannot calculate $\langle V\rangle$ by simply finding the eigenvalues of the Dirac operator. This is analogous to the famous sign problem in theories with finite chemical potential.

In our case we can provide a simple remedy to this problem. The Minkowski theory admits two distinct Wick rotations. Starting with our Minkowski formulation let us instead of eq.(\ref{wick}) 
perform the following transformation:

\be t\rightarrow ix;\ \ \ \ \ \ \phi\rightarrow i\phi\ee 
The Fermi fields transform as:

\begin{align}\begin{split}
\psi^E_1(y)&= A \exp\left({-\frac{2\pi i}{\beta}\int_{-\infty}^y d \lambda \phi'(\lambda) + \frac{1}{2}  \beta \phi(y)}\right) \equiv \xi_1 \\
(\psi^\dagger_1(y))^E&= A \exp\left({\frac{2\pi i}{\beta}\int_{-\infty}^y d \lambda \phi'(\lambda) - \frac{1}{2}  \beta \phi(y)}\right) \equiv \xi^\dagger_2\\
\psi^E_2(y)&=-i A \exp\left({-\frac{2\pi i }{\beta}\int_{-\infty}^y d \lambda \phi'(\lambda) - \frac{1}{2}  \beta \phi(y)}\right) \equiv -i\xi_2\\
(\psi^\dagger_2(y))^E&=i A \exp\left({\frac{2\pi i }{\beta}\int_{-\infty}^y d \lambda \phi'(\lambda) + \frac{1}{2}  \beta \phi(y)}\right)  \equiv i \xi^\dagger_1
\end{split}\end{align}
which leads to the Euclidean Lagrangian:

\be \mathcal{\bar L}= \xi^\dagger_2\left(\partial_x + i\partial_y\right)\xi_1 + \xi^\dagger_1\left(\partial_x-i\partial_y\right)\xi_2 +i m_1\left(\xi^\dagger_1\xi_1-\xi^\dagger_2\xi_2\right)+i m_2\left(\xi^\dagger_2\xi^\dagger_1-\xi_2\xi_1\right)  \label{yenil}\ee
and the following representation of the vortex operator
\be V(x)=\exp{\left(-i\pi\int^x (\xi^\dagger_1\xi_2 + \xi^\dagger_2\xi_1)\right)} \label{veuc}\ee
In this new representation the vortex field is real, and so is the Euclidean Lagrangian. Thus the vortex correlation functions can be calculated by calculating the eigenvalues of the Dirac operator modified by the appropriate source terms. 
 Obviously, the Polyakov operator now is complex, but this is not an issue since we already  have a convenient real Euclidean representation for $P$. 

In fact there is an interesting duality between the two representations. 
A simple transformation  $\xi_1\rightarrow i\xi_2$, $\xi_2 \rightarrow -i\xi_1$, results in $\mathcal{\bar L}(m_1,m_2)\rightarrow \mathcal{ L}(m_2,m_1)$ as well as $V\rightarrow P$.
Thus  we  learn  that we can calculate the vortex operator correlations functions and then by a simple transformation $m_1\leftrightarrow m_2$  read off the  correlation functions  of the Polyakov loop.
Our choice will be to deal directly with $V$ and the Euclidean Lagrangian eq.(\ref{yenil}) transformed by $\xi_1\rightarrow i\xi_2$, $\xi_2 \rightarrow -i\xi_1$ , and infer the results for the Polyakov loop from this calculation.

Note that it is not possible to find a representation in which $P$ and $V$ are simultaneously real. Thus were we interested in joint correlation functions we would have to deal with complex sources. Such mixed objects are  not particularly natural objects, since  the vortex field is an operator in the Hilbert space of the original theory, while Polyakov loop does not correspond to such an operator, but rather to  an auxiliary field which imposes gauge invariance\cite{chris}. Nevertheless a representation including both objects, like the one we are employing here is frequently useful in analyzing properties of finite temperature gauge theories\cite{chiral}. In this paper we are not going to be interested in such mixed correlators.

\section{Expectation values of $P$ and $V$ across the phase transition.}

Using the results of the previous section, we can represent the calculation of the expectation value of the magnetic vortex operator as the following path integral:

\be  <V> = \int D \xi D\bar{\xi} e^{-\int\mathcal{L}_V}.\ee
where

\be \mathcal{L}_V=   -\left(\xi^\dagger_2\left[\partial_x -i  \partial_y -i\pi \delta(x)\Theta(-y)\right]\xi_1 + \xi^\dagger_1\left[\partial_x +i  \partial_y - i\pi\ \delta(x)\Theta(-y)\right] \xi_2\right)
-i m_2\left(\xi^\dagger_2\xi^\dagger_1 - \xi_2\xi_1 \right)-i m_1\left( \xi^\dagger_1\xi_1-\xi^\dagger_2\xi_2\right).\label{lv} \ee

 We know from the analysis of \cite{dkkt} that the phase transition occurs when $m_1=m_2$.  In the low temperature phase, where $m_1>m_2$ the magnetic symmetry is broken and thus $\langle V\rangle\ne 0$, while in the high temperature phase the symmetry is restored and the expectation value of the vortex operator should vanish. In the current fermionic formulation the only mechanism that can make the VEV of $V$  vanish is an appearance of a normalizable zero mode in the Dirac operator. Thus our first goal is to see whether in fact the Dirac operator defined in eq.(\ref{lv}) has zero modes for $m_1<m_2$, but none for $m_1>m_2$.

\subsection{Fermionic zero modes.}  
The operator in eq.(\ref{lv}) looks singular due to appearance of the cut. This singularity however can be ``gauged away" by a unitary transformation.  
Let us therefore perform the unitary transformation - change of variables in the path integral:

\be \xi_1 \rightarrow e^{-i\frac{\theta}{2}}\xi_1, \mbox{ }\xi_2 \rightarrow e^{-i\frac{\theta}{2}}\xi_2. \label{rotation}\ee
Here the polar angle $\theta$ is defined to have exactly the same cut as before.  We treat the transformation in eq.(\ref{rotation})  as single valued but discontinuous across the cut of $\theta$.
The effect of this transformation is that the derivative of $\theta$ across the cut cancels the singular term in eq.(\ref{lv}). On the other hand the continuous part of $\partial_\mu\theta$ turns  the derivatives in eq.(\ref{lv}) into covariant derivatives in the background of a vector potential of a pointlike magnetic vortex in a regular gauge: 

\be D_\mu = \partial_\mu - \frac{i}{2}A_\mu;\ \ \ \ \ \ \  \  A_\mu=\epsilon_{\mu\nu}\frac{x_\nu}{x^2}. \ee

The Lagrangian $\mathcal{L}_V$  becomes:

\be \mathcal{L}_V = -\xi^\dagger_2\left( D_x -i D_y\right)\xi_1 -\xi^\dagger_1\left( D_x +i  D_y\right)\xi_2 -i m_2\left( e^{i\theta}\xi^\dagger_2\xi^\dagger_1 -e^{-i\theta} \xi_2\xi_1 \right)-i m_1\left( \xi^\dagger_1\xi_1-\xi^\dagger_2\xi_2\right)\label{vp.}\ee
To find possible zero modes we have to solve the equations of motion:
\begin{align}\begin{split}
\frac{\delta \mathcal{L}_V}{\delta \xi^\dagger_2} & = -\left(D_x - iD_y\right)\xi_1 - im_2e^{i\theta}\xi^\dagger_1+im_1\xi_2=0,\\
\frac{\delta \mathcal{L}_V}{\delta \xi^\dagger_1} &=-\left(D_x + iD_y\right)\xi_2 +im_2e^{i\theta}\xi^\dagger_2-im_1\xi_1=0.\\
\end{split}\end{align}

or in polar coordinates:

\begin{align}\begin{split}
e^{-i\theta}\left(i\partial_r + \frac{1}{r}\partial_\theta-\frac{i}{2r}\right)\xi_1 + im_2e^{i\theta}\xi^\dagger_1-im_1\xi_2 &=0 \label{ppol},\\
e^{i\theta}\left(-i\partial_r + \frac{1}{r}\partial_\theta-\frac{i}{2r}\right)\xi_2 -im_2e^{i\theta}\xi^\dagger_2+im_1\xi_1&=0.\\
\end{split}\end{align}
In a particular case $m_1=0$ the zero modes where found in \cite{jackiw}.
In the general case the solution can be sought in a separable form 

\be \xi_i = R_i(r)e^{i(\alpha_i+n_i\theta)}.\label{ansatz}\ee

for constant $\alpha_i$, real $R_i(r)$ and integers $n_i$. It is easy to see that the angular dependence is solved  for $n_1=1, \mbox{ }n_2=0$. The radial equations are also readily solved with the following  result. We find three independent solutions (we assume $m_1>0,\ m_2>0$ throughout):

\begin{align}\begin{split}
\xi^{asy,1}_1 &= \frac{c_1}{2\sqrt{r}}e^{-(m_2-m_1)r},\\
\xi^{asy,1}_2 &= \frac{c_1}{2\sqrt{r}} e^{-(m_2-m_1)r},\label{sol1}\\
\end{split}\end{align}

\begin{align}\begin{split}
\xi^{asy,2}_1 &=i \frac{c_2}{2\sqrt{r}}e^{-(m_1-m_2)r}e^{i\theta},\\
\xi^{asy,2}_2 &=- i\frac{c_2}{2\sqrt{r}} e^{-(m_1-m_2)r},\label{sol2}\\
\end{split}\end{align}

\begin{align}\begin{split}
\xi^{asy,3}_1 &= \frac{c_3}{2\sqrt{r}}e^{-(m_2+m_1)r}e^{i\theta},\\
\xi^{asy,3}_2 &=- \frac{c_3}{2\sqrt{r}} e^{-(m_2+m_1)r}\label{sol3}\\
\end{split}\end{align}
with real constant $c_i$. All of these solutions are normalizable in the ultraviolet. In the infrared $\xi^{asy,1}$ is normalizable for $m_2>m_1$, while $\xi^{asy,2}$ is normalizable for $m_1>m_2$ . 

 This result is somewhat surprising, since it suggests that 
there exists a normalizable zero mode for any sign of $m_1-m_2$. Translated into the language of $\langle V\rangle$ that means that  $\langle V\rangle=0$ at any temperature, while we expect it to vanish only above the phase transition. 

The resolution of this apparent puzzle is the following. It is perfectly possible for an operator to have a vanishing  VEV, even if it transforms nontrivially under a spontaneously broken symmetry. One natural reason could be a singular UV behavior. Indeed, recall that $V$ is defined as an exponential of a scalar field $\phi$, which in the UV is essentially free. The propagator of $\phi$ diverges logarithmically a short distances, and thus we indeed expect $\langle V\rangle\propto \Lambda^{-k\beta^2}\rightarrow 0$ in the continuum limit at any temperature. The existence of a normalizable zero mode in eq.(\ref{ppol}) at any temperature is the manifestation of this feature.

Analogous situation is encountered in calculation of the Polyakov loop in 3+1 gauge theories \cite{rob}. One way of dealing with it is to define a renormalized operator, since the UV divergence can indeed be canceled by  a multiplicative renormalization \cite{rob}.
We choose a different approach which is simper to implement in our framework.  In order to see whether the vanishing of $\langle V\rangle$ is related to restoration of magnetic symmetry we can modify the vortex operator by softening its UV behavior. The VEV of such modified operator then will genuinely reflect the realization of magnetic symmetry in the thermal ensemble.

\subsection{UV regularized magnetic vortex.}
To regularize the vortex operator in the UV we could go back to our original definition eq.(\ref{vo}) and smear the field $\chi$ in the exponent with some smooth test function over a finite distance scale $a$. It is simpler however to regularize directly the fermionized expression eq.(\ref{lv}).  Eq. (\ref{lv}) is just a Lagrangian of charged fermions coupled to a pointlike ``magnetic vortex" described by the vector potential $A_\mu$. Clearly,  to regulate this  expression in the UV we should smear the magnetic flux over some small area while keeping the total magnetic flux of the vortex fixed. Technically we find it the simplest to smear the flux over a ring of radius $a$ while setting the ``magnetic field" at $r<a$ to zero. This amounts to setting $A_\mu(r)=0$ for $r<a$, and leaving the vector potential at $r>a$ unmodified. 

The advantage of this procedure is that we can use the solutions eqs.(\ref{sol1},\ref{sol2}) unmodified for $r>a$ and match them at $r=a$ onto solution of the equations without ``magnetic field":
 
 \begin{align}\begin{split}
e^{-i\theta}\left(i\partial_r + \frac{1}{r}\partial_\theta \right)\xi_1 + im_2e^{i\theta}\xi^\dagger_1-im_1\xi_2 &=0, \\
e^{i\theta}\left(-i\partial_r + \frac{1}{r}\partial_\theta\right)\xi_2 -im_2e^{i\theta}\xi^\dagger_2+im_1\xi_1&=0.\label{eqreg}\\
\end{split}\end{align}

To solve eq.(\ref{eqreg}), we again look for solutions of the form eq.(\ref{ansatz}). By inspection, we see that the only allowed  angular dependence is:

\begin{align}\begin{split}
\xi_1 &= R_1(r)e^{i\alpha_1}e^{i\theta},\\
\xi_2 &= R_2(r)e^{i\alpha_2}.\\
\end{split}\end{align}

The general solution is a superposition of the following  independent solutions with {\it real} coefficients, where $I$ and $Y$ are modified Bessel functions:

\begin{align}\begin{split}
&\xi^{int,1}_1=  e^{-m_2 r} \left(d_1 I_1(m_1 r)+i d_2 Y_1(-i m_1 r)\right)e^{i\theta},\\
& \xi^{int,1}_2= e^{-m_2 r} \left(d_1 I_0( m_1 r)+d_2 Y_0(-i m_1 r)\right),\\
\end{split}\end{align}

\begin{align}\begin{split}
&\xi^{int,2}_1=  e^{m_2 r} \left(i d_3 I_1(m_1 r)- d_4 Y_1(-i m_1 r)\right)e^{i\theta},\\
& \xi^{int,2}_2= e^{m_2 r} \left(i d_3 I_0( m_1 r)+i d_4 Y_0(-i m_1 r)\right),\\
\end{split}\end{align}

 with real constants $d_i$. Since $Y_1(r)\rightarrow 1/r$ at $r\rightarrow 0$, normalizability  at $r\rightarrow 0$ requires $d_2=d_4=0$. Thus the most general solution for $r<a$ is:

\begin{align}\begin{split}
&\xi^{int}_1=  \left(\alpha e^{-m_2 r}+ i \beta e^{m_2 r}\right)I_{1}(m_1 r)e^{i\theta},\\
& \xi^{int}_2= \left(\alpha e^{-m_2 r}+ i \beta e^{m_2 r}\right)I_{0}(m_1 r)\label{ic}.\\
\end{split}\end{align}
This should match a linear combination of eqs.(\ref{sol1}),( \ref{sol2}) and (\ref{sol3}) at $r=a$.

The solution eq.(\ref{sol1}) is normalizable for $m_2> m_1$, while eq.(\ref{sol2}) is normalizable for  $m_1>m_2$. 
Consider first the low temperature phase $m_1>m_2$. In this case the most general solution for $r>a$  is:

\begin{align}\begin{split}
\xi^{asy}_1 &=\frac{1}{2\sqrt{r}}e^{-m_1 r}\left( A e^{-m_2 r}+ i B e^{m_2 r} \right)e^{i\theta},\\
\xi^{asy}_2 &=\frac{1}{2\sqrt{r}}e^{-m_1 r}\left(- A e^{-m_2 r}- i B e^{m_2 r} \right)\label{dis}.\\
\end{split}\end{align}

Continuity of eqs.(\ref{ic}) and (\ref{dis}) at $r=a$ requires:

\begin{align}\begin{split}
 \frac{A}{\alpha} &= \frac{B}{\beta} = I_1 (m_1 a)e^{m_1 a} 2 \sqrt{a},\\
  \frac{A}{\alpha} &= \frac{B}{\beta} = - I_0 (m_1 a)e^{m_1 a} 2 \sqrt{a},\\
\end{split}\end{align}

These two conditions cannot  be satisfied simultaneously since the modified Bessel functions are always positive for all $a>0$. It thus follows that for  $m_1>m_2$ there are no normalizable zero modes. This means that the UV nonsingular vortex operator has a nonvanishing expectation value in the low temperature phase, consistent with our expectation.

In the high temperature phase, for $m_2>m_1$  the exterior solution is:

\begin{align}\begin{split}
\xi^{asy}_1 &= \frac{1}{2\sqrt{r}}\left(c_1e^{-(m_2-m_1)r}+ c_2 e^{-(m_2+m_1)r}\right)e^{i\theta},\\
\xi^{asy}_2 &= \frac{1}{2\sqrt{r}}\left(c_1 e^{-(m_2-m_1)r}- c_2 e^{-(m_2+m_1)r}\right).\\
\end{split}\end{align}

 This time the matching across $r=a$ requires

\begin{align}\begin{split}
\alpha I_1(m_1 a) &= \frac{1}{2\sqrt{a}}\left(c_1 e^{m_1 a} + c_2 e^{-m_1 a}\right),\\
\alpha I_0(m_1 a) &= \frac{1}{2\sqrt{a}}\left(c_1e^{m_1 a} - c_2 e^{-m_1 a}\right),
\end{split}\end{align}
or since $m_1a\ll 1$:

\begin{align}\begin{split}
0&= \frac{1}{2\sqrt{a}}\left(c_1 + c_2 \right),\\
\alpha  &= \frac{1}{2\sqrt{a}}\left(c_1 - c_2 \right),
\end{split}\end{align}

These conditions can obviously be satisfied:
\be c_1= -c_2= \alpha a^{1/2}\ee
The only acceptable solution for $T>T_c$ is (for $r>a$)
\begin{align}\begin{split}
\xi^{asy}_1 &= \frac{A}{2\sqrt{r}}\left(e^{-(m_2-m_1)r}- e^{-(m_2+m_1)r}\right)e^{i\theta},\\
\xi^{asy}_2 &= \frac{A}{2\sqrt{r}}\left(e^{-(m_2-m_1)r}+ e^{-(m_2+m_1)r}\right).\\
\end{split}\end{align}

Thus in the high temperature phase we find that the normalizable zero mode survives, and therefore $\langle V\rangle=0$.

So far we have studied the zero modes in the calculation of the vortex operator. As explained in the previous section, this  is sufficient to obtain equivalent information about the Polyakov loop. We have seen that the path integral for calculation of $\langle P\rangle$ is identical to that for calculation of $\langle V\rangle$ with the sole substitution  $m_1 \leftrightarrow m_2$. Thus the normalizable zero mode exist in this case for $m_1>m_2$, and we confirm that $\langle P\rangle=0$ below $T_c$ and $\langle P\rangle\ne 0$ for $T>T_c$.

Although the results of this section are not surprising, this is the first calculation that we are aware of which directly reproduces the  behavior of the deconfinement order parameter(s) in the vicinity of the phase transition using the fermionized version of the partition function. In the next section we will use the same representation to calculate the correlation function of the vortex and Polyakov operators.

\section{Correlators of the Vortex Operator}
Our next goal is to calculate the correlation function of the vortex operators. There are two distinct correlation functions  of interest: $\langle V(x)V^*(y)\rangle$ and $\langle V(x)V(y)\rangle$. In this section we will calculate the large distance behavior of these correlators.

\subsection{The Roadmap to the Calculation}

Th correlation function of a vortex located at the origin and an antivortex at a point $\vec l$ is given by the path integral 

\begin{align}
\braket{V(0)V^*(l)}=\int D\xi D\bar{\xi} e^{\int \mathcal{L}_{VV^*}},
\end{align}
with
\begin{eqnarray} \mathcal{L}_{VV^*}&=&   -\left(\xi^\dagger_2\left[\partial_x -i  \partial_y \right]\xi_1+ \xi^\dagger_1\left[\partial_x +i  \partial_y\right]\xi_2\right)
+i\pi\left[\delta(x)\Theta(-y)-\delta(x-l_1)\Theta(l_2-y)\right]\left(\xi^\dagger_2\xi_1 +\xi^\dagger_1\xi_2\right) \nonumber\\
&-&i m_2\left(\xi^\dagger_2\xi^\dagger_1 - \xi_2\xi_1 \right)-i m_1\left( \xi^\dagger_1\xi_1-\xi^\dagger_2\xi_2\right).\label{lvv^*} \end{eqnarray}

The cut discontinuity again can be removed by a rotation of the spinors similar to that in  eq.(\ref{rotation}):

\be \xi_1 \rightarrow e^{-i\frac{\theta-\theta'}{2}}\xi_1, \mbox{ }\xi_2 \rightarrow e^{-i\frac{\theta-\theta'}{2}}\xi_2 \label{vvstar}.\ee

Here the angle $\theta$ is defined in eq.(\ref{angledef}) , and $\theta'$ is the angle between the vector $\vec x-\vec l$ and the direction of the cut (the negative $y$ axis). (From this point on, we will use primed characters for objects defined with respect to the antivortex location $\vec l$)
The resulting expression is
\be \mathcal{L}_{VV^*} = -\xi^\dagger_2\left( D_x -i D_y\right)\xi_1 -\xi^\dagger_1\left( D_x +i  D_y\right)\xi_2 -i m_2\left( e^{i(\theta-\theta')}\xi^\dagger_2\xi^\dagger_1 -e^{-i(\theta-\theta')} \xi_2\xi_1 \right)-i m_1\left( \xi^\dagger_1\xi_1-\xi^\dagger_2\xi_2\right) \label{vvstarl},\ee

\be D_\mu = \partial_\mu - \frac{i}{2}(A_\mu-A'_\mu) .\ee

Our aim is to calculate the correlator for the case where the separation between the vortex and the antivortex is larger than the intrinsic length scales in this theory ($l\gg m^{-1}_1,m^{-1}_2$). Thus we follow the standard logic employed in such cases, see e.g. \cite{shuryak}.
One expect that most of the eigenvalues of the operator in eq.(\ref{vvstarl}) are of the order of the mass scale in the theory ($\sim m_1,m_2$), and therefore do not depend on $l$ in the interesting regime. The exception are the two eigenvalues that correspond to zero eigenvalues in the limit $l\rightarrow\infty$. The leading $l$ dependence is therefore given by the subdeterminant on the space of functions spanned by these quasi zero modes. We will therefore concentrate on the subspace of functions spanned by the vortex and antivortex zero modes, and will calculate the subdeterminant on this subspace. The leading $l$ dependence is given by this factor in the full determinant, whereas the rest of the determinant only contributes to the overall normalization of the correlator. 

Since our operator eq.(\ref{vvstarl}) has a term which mixes $\xi^\dagger$ with $\xi^\dagger$ we must be a little careful with our calculation. The proper way of proceeding is to represent the path integral in the form

\begin{align}
\int D\theta D\sigma e^{\int v^T L_{VV^*} v}
\end{align}

where $v^T=\begin{pmatrix}\theta_1&\theta_2&\sigma_1&\sigma_2\end{pmatrix}$, $\theta$ and $\sigma$ are the real and imaginary parts of the spinor $\xi$, and $L_{VV^*}$ is explicitly antisymmetric. We can  then formally complexify $\theta,\sigma$ and perform the path integral:

\begin{align}\label{zj}
\int D\theta D\sigma e^{\int v^T {L}_{VV^*} v}=(\det{L}_{VV^*})^{1/2}
\end{align}
with the operator $L_{VV^*}$ 

\begin{align}\begin{split}
L_{VV^*} = \begin{pmatrix} 0 & -\partial_x + \frac{\Delta A_y}{2}+m_2\sin{\Delta\theta} & - m_1 &  \partial_y + \frac{\Delta A_x}{2}-m_2\cos{\Delta\theta}  \\
			 -\partial_x - \frac{\Delta A_y}{2}-m_2\sin{\Delta\theta} & 0 &  -\partial_y + \frac{\Delta A_x}{2}+m_2\cos{\Delta\theta} & m_1 \\
			m_1 &  -\partial_y - \frac{\Delta A_x}{2}-m_2\cos{\Delta\theta} & 0 &  -\partial_x + \frac{\Delta A_y}{2}-m_2\sin{\Delta\theta} \\
			 \partial_y - \frac{\Delta A_x}{2}+m_2\cos{\Delta\theta} & -m_1 &  -\partial_x - \frac{\Delta A_y}{2}+m_2\sin{\Delta\theta} & 0 \end{pmatrix}. \label{oper}
			\end{split}\end{align}
			
where $\Delta\theta\equiv \theta-\theta'$

To calculate the eigenfunctions we first have to choose a good basis for the reduced Hilbert space. Naively we would just take the space spanned by the zero mode of the vortex and the zero mode of the antivortex. However the presence of an antivortex changes the equation of motion for the vortex zero mode since the extra phase factor multiplying $m_2$ is not small even if the antivortex is far away. In the quasi zero mode we have to account for this phase change. We choose as our basis vectors the solutions of the following equations\footnote{We are being somewhat schematic in our notations here, but the exact form of the equation should be clear from the context.}
\begin{align}\label{s8}
(\partial-\frac{i}{2}A)\psi-im_2e^{i(\theta-\theta'_0)}\psi^\dag+im_1\psi=0, 
\end{align}
\begin{align}\label{s9}
(\partial+\frac{i}{2}A')\eta-im_2e^{i(\theta_0-\theta')}\eta^\dag+im_1\eta=0, 
\end{align}
where $\theta'_0$ is the value of $\theta'$ at the location of the antivortex.  The logic here is that the solution $\psi$ as we know decays exponentially away from the center of the vortex. Thus the extra angular factor $e^{i\theta'}$ that multiplies $m_2$ equals to $e^{i\theta'_0}$ in the part of plane where the function $\psi$ is significantly different from zero. The same argument holds for the angular factor in eq.(\ref{s9}).
 
Given the solutions of eqs.(\ref{s8},\ref{s9}) we have to calculate the reduced matrix
\begin{align}
\bra{\eta}L_{VV^*}\ket{\eta} ,\quad \bra{\psi}L_{VV^*}\ket{\psi},\quad \bra{\eta}L_{VV^*}\ket{\psi}, \quad \bra{\psi}L_{VV^*}\ket{\eta}.
\end{align}
This calculation is simplified by the fact that the basis vectors $\psi$ and $\eta$ correspond to real functions $\theta$ and $\sigma$. Since $L_{VV^*}$ is a real antisymmetric operator, its diagonal matrix elements vanish. Thus the only matrix element that has to be calculated is 

\begin{align*}
\bra{\psi}L_{VV^*}\ket{\eta}\equiv \lambda_1
\end{align*}

%An eigenmode of this system satisfies 

%\begin{align}\label{s3}
%(\partial+A-A')\xi-im_2e^{i(\theta-\theta')}\xi^\dag+im_1\xi=\lambda \xi.
%\end{align}

%To find the zero modes of this system we can do the following. Due to large separation between sources,
%the lowest energy state of the system can be expressed as 

%\begin{align}
%\ket{\psi^0_{12}}=\ket{\psi^0_1}\ket{\psi^0_2},
%\end{align} 

%where $\psi^0_i(x)$ are zero modes which are localized around each source, $\eta$ at the anti-vortex and $\psi$ at the vortex. In this case $\psi$ must satisfy

%After plugging $\psi$ in \refeq{s3} and subtracting this zero contribution we get

%\begin{align}
%-A'\psi-im_2\Big[ e^{i(\theta-\theta')}-e^{i(\theta-\theta'_0)} \Big]\psi^\dag=\lambda_1\psi,
%\end{align}

Thus the relevant sub matrix is

\begin{align}
\mathcal{L}_{VV^*}|_{\eta,\psi}=\begin{pmatrix}0&\lambda_1\\-\lambda_1&0\end{pmatrix}.
\end{align}

From which we deduce, following eq.(\refeq{zj})

\begin{align}
\braket{VV^*}\approx\lambda_1.
\end{align}

Our aim in the next subsection is to calculate the matrix element $\lambda_1$.

\subsection{The quasi zero modes.}

The quasi zero mode for the vortex (in the presence of the antivortex) is given by the solution of 

\begin{align}\begin{split}
e^{i\theta}\left[i\partial_r - \frac{1}{r}\partial_\theta + \frac{i}{2r}\right]\psi_2 + i m_2 e^{i(\theta-\theta'_0)}\psi_2^*-im_1\psi_1&=0,\\
 e^{-i\theta}\left[-i\partial_r - \frac{1}{r}\partial_\theta + \frac{i}{2r}\right]\psi_1 - i m_2 e^{i\theta-\theta'_0)}\psi_1^*+im_1\psi_2&=0. \label{vortexeq}
\end{split}\end{align}

Comparing with eq.(\ref{ppol}) we see that the only difference is the extra  phase multiplying $m_2$. It  can be accounted for easily by an additional global phase rotation. 
The solution therefore is:

\begin{align}\begin{split}
\psi_1&= \frac{A}{\sqrt r}\left[e^{-(m_2-m_1)r}-e^{-(m_1+m_2)r}\right]e^{i(\theta-\frac{\theta'_0}{2})},\\
\psi_2&=  \frac{A}{\sqrt r}\left[e^{-(m_2-m_1)r}+e^{-(m_1+m_2)r}\right]e^{-i\frac{\theta'_0}{2}}, \\ \label{zerovv}
\end{split}\end{align}

The antivortex quasi zero mode satisfies:

\begin{align}\begin{split}
e^{i\theta''}\left[i\partial_{r'} - \frac{1}{r'}\partial_{\theta'} - \frac{i}{2r'}\right]\eta_2 +i m_2 e^{-i(\theta'-\theta_0)}\eta_2^* - im_1 \eta_1&=0,\\
e^{-i\theta'}\left[-i\partial_{r'} - \frac{1}{r'}\partial_{\theta'} - \frac{i}{2r'}\right]\eta_1 -i m_2 e^{-i(\theta'-\theta_0)\theta''_0}\eta_1^* + im_1 \eta_2&=0,\label{zeroav}
\end{split}\end{align}

where $r'$ is the distance from the location of the antivortex . Eq.(\ref{zeroav}) is related  to eq.( \ref{vortexeq}) by a simple transformation: 
\be \psi_{1,2 }(r,\theta) \rightarrow\eta^*_{2,1}(r',\theta'). \ee

% \begin{figure}[t]
%
%\begin{tikzpicture}[thick]
%\coordinate (O) at (0,0);
%\coordinate (K) at (2.4,-2.7);
%\coordinate (L) at (0,-5);
%\coordinate (M) at (5,-5);
%\coordinate (N) at (5,2);
%\draw (L)--(O)--(N)--(M);
%\draw [dashed] (O)--(K)--(N);
%\draw [ultra thick, ->-=0.5] (O)--(L);
%\draw [ultra thick,->-=0.5 ] (M)--(N);
%\foreach \pt/\labpos in {O/above,K/right,N/above left}{
%      \filldraw (\pt) circle(.8mm) node[\labpos=1.5mm,fill=white]{\pt};
%    }
%
%
%\draw [dashed]  (O)--(K) node[pos=0.5,,above,yshift=0.1cm,xshift=0.2cm] {\LARGE $r$};
%\draw [dashed] (K)--(N) node[pos=0.5,,above,xshift=-0.2cm] {\LARGE $r '$};
%\draw  (O)--(N)  ;
%
%\pic["$\theta$", draw, ->, angle eccentricity=1.2, angle radius=1cm]
%    {angle=L--O--K};
%
%\pic["$\theta'$", draw, ->, angle eccentricity=1.2, angle radius=1.6cm]
%    {angle=M--N--K};    
%\pic["$-\Delta\theta$", draw, ->, angle eccentricity=1.2, angle radius=0.6cm]
%    {angle=O--K--N};        
%2
%\end{tikzpicture}
%\caption{VV* Configuration. \label{vv*}} 
%\end{figure}
\begin{figure}[h]
	\caption{VV* Configuration. \label{vv*}}
	\includegraphics[scale=0.8]{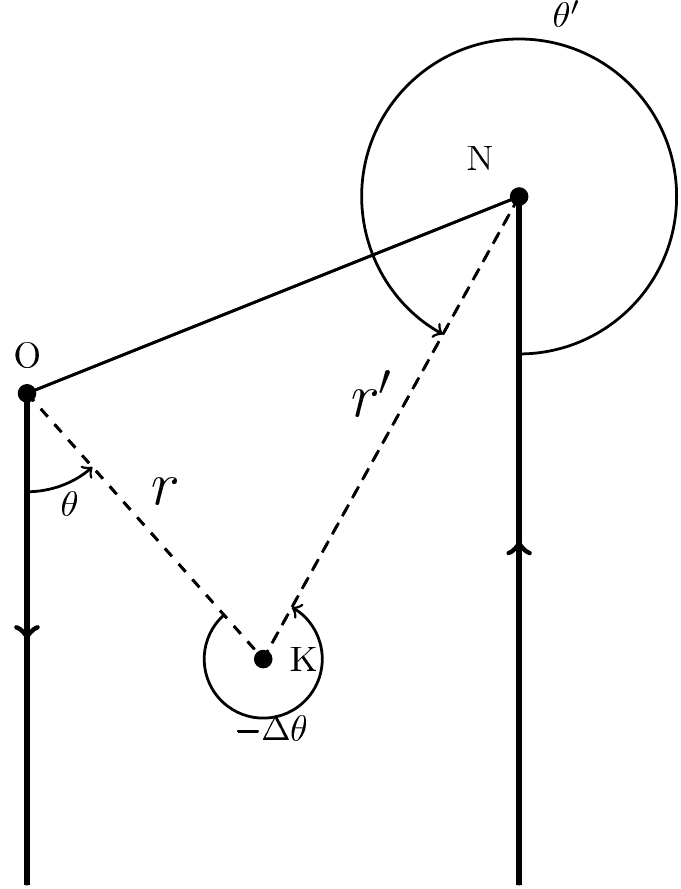}
\end{figure}
So, the antivortex quasi zero mode is:

\begin{align}\begin{split}
\eta_1&=  \frac{A}{\sqrt {r'}}\left[e^{-(m_2-m_1)r'}+e^{-(m_1+m_2)r'}\right] e^{i\frac{\theta_0}{2}},\\
\eta_2&=  \frac{A}{\sqrt{ r'}}\left[e^{-(m_2-m_1)r'}-e^{-(m_1+m_2)r'}\right]e^{-i(\theta'-\frac{\theta_0}{2})}. \\ \label{zerovs}
\end{split}\end{align}

In the following we are interested in the leading large distance behavior and will therefore disregard the subleading exponents in eqs.(\ref{zerovv},\ref{zerovs}).

\subsection{The correlator.}

Using our chosen basis for the quasi zero modes we find for the off diagonal matrix element of $L_{VV^*}$ the following expression:

\begin{align}\begin{split}
&<\eta(r')|L_{VV^*}|\psi(r)>\\
=&\int d^2x \frac{e^{-m(r+r')}}{\sqrt{rr'}}\bigg[m_2\big(1-\cos(\theta-\theta') -\cos(\theta-\theta_0)-\cos(\theta'-\theta'_0)\big)+\frac{1}{2r'}\left[\cos(\theta'-\theta'_0)-\cos(\theta-\theta_0)\right]\bigg].\label{vvscorr}\\
\end{split}\end{align}
where
\be m\equiv m_2-m_1\ee
An important feature of this expression is that it depends only on differences of the angles. This means that it is insensitive to the direction of the cut in the function $\theta$. This is an important consistency check on our calculation.

This integral can be calculated expanding in the inverse powers of $ml$. To do this we note that the main contribution to the integral in the second term in eq.(\ref{vvscorr}) comes from the vicinity of the point $r=l,\ \theta-\theta_0$.  To leading order we set $r=l$ and $\theta=\theta_0$, and obtain for the integral in eq.(\ref{vvscorr})
\begin{align}\begin{split}
&<\eta(r')|L_{VV^*}|\psi(r)>_0\\
=&\int d^2x \frac{e^{-m(l+r')}}{\sqrt{lr'}}\bigg[\frac{1}{2r'}\left[\cos(\theta'-\theta'_0)-1\right]\bigg]=-mle^{-ml}\left(\frac{\pi}{ml}\right)^{3/2}\label{vvscorr0}\\
\end{split}\end{align}

The leading contribution to the first term in eq.(\ref{vvscorr}) comes from two regions 
\be (r\approx l; \ \theta=\theta_0); \ \ {\rm and} \ (r'\approx l;\ \theta'=\theta'_0)\ee
The putative leading order term (simply setting $r=l,\ \theta=\theta_0$, etc.) vanishes. To find 
the first nonvanishing contribution we expand $\theta(\theta')$ around $\theta_0(\theta'_0)$ using the relation

\be 
\frac{l}{\sin(\theta-\theta')} = \frac{r}{\sin(\theta'-\theta'_0)}=\frac{r'}{\sin(\theta_0-\theta)}.
\ee

Define $\epsilon \equiv\theta_0-\theta$. Expanding for small $\epsilon$:

\begin{align}
\epsilon \approx \frac{r'}{l}\sin(\theta_0-\theta') = \frac{r'}{l}\sin(\theta'-\theta'_0).
\end{align}

Using this, and $\theta'_0=\theta_0+\pi$  we get

\be
\big(1-\cos(\theta-\theta') -\cos(\theta-\theta_0)-\cos(\theta'-\theta'_0)\big) \approx -\frac{r'}{l}\sin^2(\theta'-\theta'_0).
\ee

Using $\cos2\theta=1-2\sin^2\theta$ and integrating, we get for the $m_2$ term 

\begin{align}\begin{split}
&2\int d^2x \frac{e^{-m(r+r')}}{\sqrt{lr'}}\frac{m_2 r'}{2 l} = \frac{3m_2}{m}\left(\frac{\pi}{m l}\right)^{3/2}e^{-ml}.\\
\end{split}\end{align}
where the factor 2 takes into account the contribution of the region $r'\approx l$. Although this term has one factor of $l$ less than eq.(\ref{vvscorr0}), we keep it since the ratio $m_2/m$ diverges at the phase transition.

In all we obtain
\be
<\eta(r')|L_{VV^*}|\psi(r)>=-
\left[3\frac{m_2}{m}+ml\right]e^{-ml}\left(\frac{\pi}{ml}\right)^{3/2}\label{vvscorrf}
\ee

%\begin{align}\begin{split}
%e^{-ml}\left(\frac{\pi}{ml}\right)^{3/2} \left[\frac{1}{2} -ml +\frac{3}{16ml}-\frac{15}{32m^2l^2}\right]
%\end{split}\end{align}
To determine the vortex-antivortex correlator we have to determine the sign of the square root of the determinant. We do not have an unambiguous way of doing it within our current procedure. However we can compare our result with the high temperature calculation for the same correlator in \cite{hight}. By continuity we believe that the sign for large $l$ should be the same as in \cite{hight}. This determines the correlator as
\be \langle V(0)V^*(l)\rangle=\left[ml+3\frac{m_2}{m}\right]e^{-ml}\left(\frac{\pi}{ml}\right)^{3/2}\label{vvscorrff}
\ee

\section{Scalar vs pseudoscalar correlations.}
It is interesting to compare the correlations in the scalar and pseudoscalar channels.  We define the scalar and pseudoscalar fields
\be S(x)\equiv  \frac{1}{\sqrt 2}\left[V(x)+V^*(x)\right];\ \ \ \ \ \ P(x)\equiv \frac{i}{\sqrt 2}\left[V(x)-V^*(x)\right]\ee

The question here is somewhat analogous to the question about approximate restoration of the $U_A(1)$ axial symmetry in QCD. Naively one expects that at high temperature instanton effects become unimportant. Since the splitting between the scalar and pseudoscalar correlation lengths is due to instantons, one may expect that the anomalous $U_A(1)$ symmetry is restored at high $T$ and the two channels become degenerate. 

Similarly, in the present model the magnetic symmetry of the Georgi-Glashow model is anomalously broken from $U(1)$ to $Z_2$ by monopole effects. If the magnetic symmetry was $U(1)$, above the transition the vortex-vortex correlator would vanish, and the scalar and pseudoscalar correlators would be identical.  

In  \cite{hight} these correlation functions were calculated at high temperatures. Although the $VV$ correlator does not vanish at high temperature, it is strongly supressed relative to the $VV^*$ correlator. The scalar and pseudoscalar correlation lengths are split, but the splitting  is very small at high temperatures. Similar behavior is observed in QCD in the ``instanton chain" approximation \cite{dk}.

In the present paper we are in a position to calculate the correlators in question close to critical temperature. We expect that the breaking of the magnetic $U(1)$ symmetry close to critical temperature is an order one effect.

\subsection{The $<VV>$ correlator. \label{here}}
The calculation of this correlator proceeds in a very similar fashion to that of $\langle VV^*\rangle$. 
The only difference is that the two elements of the basis are both quasi zero modes of a vortex, $\psi(r)$ and $\psi(r')$.
As a result for the off diagonal matrix element we have

\begin{equation}
<\psi(r')|L_{VV}|\psi(r)>
=\int d^2x \frac{e^{-m(r+r')}}{\sqrt{rr'}}\bigg[m_2\big(1-\cos(\theta-\theta') -\cos(\theta-\theta_0)-\cos(\theta'-\theta'_0)\big)- \frac{1}{2r'}\left[1+\cos(\theta-\theta')\right]\bigg].\label{vvcor}\\
\end{equation}

This can be written as
\be <\psi(r')|L_{VV}|\psi(r)>=<\eta(r')|L_{VV^*}|\psi(r)>+L_-\ee
with

\be
L_-=-\int d^2x \frac{e^{-m(r+r')}}{\sqrt{rr'}}\frac{1}{2r'}\bigg[1 + \cos(\theta-\theta') +\cos(\theta'-\theta'_0)-\cos(\theta-\theta_0)\bigg]
\ee
Using the same approximation as in the previous section we find to leading order in $1/ml$

\be L_-=-\frac{5}{4}e^{-ml}\left(\frac{\pi}{ml}\right)^{3/2} \ee
Then
\be \langle V(0)V(l)\rangle=\langle V(0)V^*(l)\rangle -L_-= \left[ml+3\frac{m_2}{m}  +\frac{5}{4}\right]e^{-ml}\left(\frac{\pi}{ml}\right)^{3/2} \ee

\subsection{Scalar vs  Pseudoscalar.}

Given the vortex and antivortex correlators, we find 
\be \langle S(0)S(l)\rangle=2\left[ml+3\frac{m_2}{m} \right]e^{-ml}\left(\frac{\pi}{ml}\right)^{3/2}\ee
\be \langle P(0)P(l)\rangle=-\frac{5}{4}e^{-ml}\left(\frac{\pi}{ml}\right)^{3/2}\ee
It is instructive to compare these results with the high temperature expressions calculated in \cite{hight}. At high temperature
\be 
\langle V(0)V(l)\rangle\ll  \langle V(0)V^*(l)\rangle,\ee
and consequently

\be  \langle S(0)S(l)\rangle\propto e^{-l\tilde\sigma_+}; \ \ \ \  \langle P(0)P(l)\rangle\propto e^{-l\tilde\sigma_-} \ee
where
\be \frac{\sigma_--\sigma_+}{\sigma_-+\sigma_+}=e^{-\kappa\frac{\pi}{g^2}}\ee
with $\kappa$ -  a number of order one.
 
 Our present results show that close to criticality 
 
 \be 
\langle V(0)V(l)\rangle\approx \langle V(0)V^*(l)\rangle.\ee 
Thus indeed the breaking of the $U(1)$ magnetic symmetry is an order one effect and is not at all suppressed close to criticality. When expressed in terms of the scalar and pseudoscalar correlators, it means,
 \be \label{ineq}
\langle S(0)S(l)\rangle\gg \langle P(0)P(l)\rangle.\ee  
This is indeed what we see in our results.
Usually the inequality eq.(\ref{ineq})  is interpreted to mean that the scalar correlation length is larger than the pseudoscalar one. Interestingly this is not the case in our calculation. We do not see splitting between the scalar and pseudoscalar correlation lengths at all. Instead, the smallness of the pseudoscalar correlator is due to a faster decay of a power prefactor rather than a smaller correlation length.

It is possible that splitting between the scalar and pseudoscalar correlation lengths is a higher order effect in the four fermi coupling which we have neglected in this calculation.

 \section{Thermodynamic quantities.}
The last question we address in this paper is calculation of entropy and free energy close to criticality. Our interest in these quantities stems from the properties of analogous quantities in QCD. It has long been known \cite{karsch} that in QCD both the free energy and the entropy are lower than their values for free massless gluon gas well above the critical temperature. In particular at temperatures of several $T_c$ the discrepancy is of order of 20\%. One possible source of the discrepancy is the existence of the topological defects  - monopoles or vortices in the thermal ensemble. It is thus interesting to see what is the effect of such objects in the Georgi-Glashow model, where we have analytic calculational control over proceedings.

We first ask what limit is analogous to the perturbative limit of free massless gluon gas. Thermal perturbation theory in the Georgi-Glashow model would not include the contributions of monopoles, which lead to appearance of non vanishing $m_1$. On the other hand it would contain the contribution to the Debye mass of the photon due to heavy charged bosons. This Debye mass is obviously proportional to $m_2$.  We will therefore compare the entropy and free energy of the model  for non vanishing and vanishing values of $m_1$, i.e. $\Delta F(m_1)\equiv F(m_1)-F(m_1=0)$ and $\Delta S(m_1)\equiv S(m_1)-S(m_1=0)$. 

In particular it is interesting to see if the presence of massive vortices  increases or decreases the entropy relative to the perturbative value. There are two competing effects associated with the vortices. On the one hand there is a positive entropy configurational entropy  associated with their distribution in real space, while on the other the magnetic field associated with them may carry an ordered structure, so that one might expect some entropy reduction.

To calculate the free energy of our system we consider the partition function of the fermionized model:
\begin{align}\label{z}
Z=\int D\theta D\sigma e^{\int v^T A v}=(det(A))^{1/2}
\end{align}

where $\{\theta,\sigma\}$ are real Grassmanians such that $\psi=\theta+i\sigma$, and the phase of the exponent is $-S=\int d^2 x \LL$:
\begin{align}\label{a}
\LL=\begin{pmatrix}\theta_1&\theta_2&\sigma_1&\sigma_2\end{pmatrix}
\begin{pmatrix} 0 & -\del_x & m_1 & \del_y+m_2 \\ -\del_x & 0 & -\del_y-m_2 & -m_1 \\ -m_1 & -\del_y+m_2 & 0 & -\del_x \\ \del_y-m_2 & m_1 & -\del_x & 0 \end{pmatrix}
\begin{pmatrix} \theta_1 \\ \theta_2 \\ \sigma_1 \\ \sigma_2  \end{pmatrix}
\end{align}

\subsection{The  vortex contribution to entropy.}

Naively one is tempted to associated the free energy of the model with the partition function $Z$ of eq.(\ref{z})  via the standard relation
\begin{align}\label{barf}
\bar F=-\frac{1}{\beta} lnZ
\end{align}

where $\beta=1/T$. This is however not quite right. We know that local observables in the thermal ensemble of the original model are related by bosonization relations to local observables in the fermionic theory. The bosonization relations however do not extend to the value of partition function itself in  the naive form of eq.(\ref{barf}). It is nevertheless possible to calculate $\Delta F$ using fermionic formulation. To do that we note that
\be
\frac{\partial}{\partial m_1} F(m_1)\propto \langle V^2+V^{*2}\rangle
\ee
were the partial derivative is defined at fixed $m_2$ and $T$.
Since $V^2$ is a local operator, its expectation value falls under the jurisdiction of bosonization rules. Thus
\be \Delta F(m_1)=\int_0^{m_1}\frac{\partial}{\partial m_1}\bar F(m_1)
\ee
 Using (\refeq{z}), and (\refeq{a}) we get
\begin{align}
\bar F=-TV\frac{1}{2}\int \frac{d^2p}{(2\pi)^2}\Big\{\ln(p^2+m_+^2)+\ln(p^2+m_-^2)\Big\}
\end{align}

where $m_\mp=m_1\mp m_2$, and V is the spatial volume of the system. 

The derivative of free energy with respect to vortex coupling is straightforward

\begin{align}\label{fm1}
\begin{split}
\frac{\del\bar  F}{\del m_1}&=-\frac{TV}{2}\frac{1}{(2\pi)^2}(2\pi)\int_0^\Lambda \frac{dp^2}{2}\Big\{\frac{2(m_1+m_2)}{p^2+(m_1+m_2)^2}+\frac{2(m_1-m_2)}{p^2+(m_1-m_2)^2} \Big\} \\
&=TV\frac{1}{2\pi}\Big\{ (m_1-m_2) \ln |m_1-m_2|+(m_1+m_2) \ln |m_1+m_2|-2m_1\ln T  \Big\}
\end{split}
\end{align}

where we used the fact that the dimensionally reduced theory is defined with a UV cutoff $\Lambda \sim T$ and $m_\pm/T<<1$ in all the interesting range of temperatures.
Integrating over $m_1$ we find 

\begin{eqnarray}\label{Fm1}
\Delta F(m_1)&=&\frac{TV}{4\pi}\Big\{ (m_1-m_2)^2\ln|m_1-m_2|+(m_1+m_2)^2\ln|m_1+m_2|-2m_2^2\ln(m_2)-2m_1^2\ln(T)-m_1^2  \Big\}\nonumber\\
&=&\frac{TV}{4\pi}\Big\{m_1^2\ln\frac{|m_1^2-m_2^2|}{T^2}+m_2^2\ln\frac{|m_1^2-m_2^2|}{m_2^2}+2m_1m_2\ln\frac{m_1+m_2}{|m_1-m_2|}-m_1^2\Big\}
\end{eqnarray}
This expression is logarithmically dominated by the first term, which is negative in the whole region of parameters where our calculation is valid.  Interestingly, this is the same trend as in 3+1 dimensional Yang-Mills theory \cite{karsch}. 

The vortex contribution to entropy is given by 
\begin{equation}
\Delta S(m_1)=-\frac{d}{dT}\Delta F(m_1)
\end{equation}
where we have to take into account that the mass parameters depend on temperature via
\begin{align}\label{m}
m_1\propto \frac{e^{-4\pi M_W/g^2}}{T} && m_2\propto\frac{ e^{-M_W/T}}{T} 
\end{align}
Discarding the small terms (of relative order $T/M_W$) we find:
\begin{equation}\label{sm1}
\Delta S(m_1)=\frac{V M_W m_1^2}{2\pi T}k\Big\{ (k-1)\ln\Big|1-\frac{1}{k}\Big|+(k+1)  \ln \Big(1+\frac{1}{k}\Big) \Big\}
\end{equation}
where $k=m_2/m_1$. The plot of  $\Delta S(m_1)$ as a function of $T/T_c$  is shown in Fig.1

\begin{figure}[h]
	\caption{$\Delta S(m_1)/VM_We^{-8\pi M_W/g^2}$ as a function of $T/T_c$}
	\includegraphics[scale=0.8]{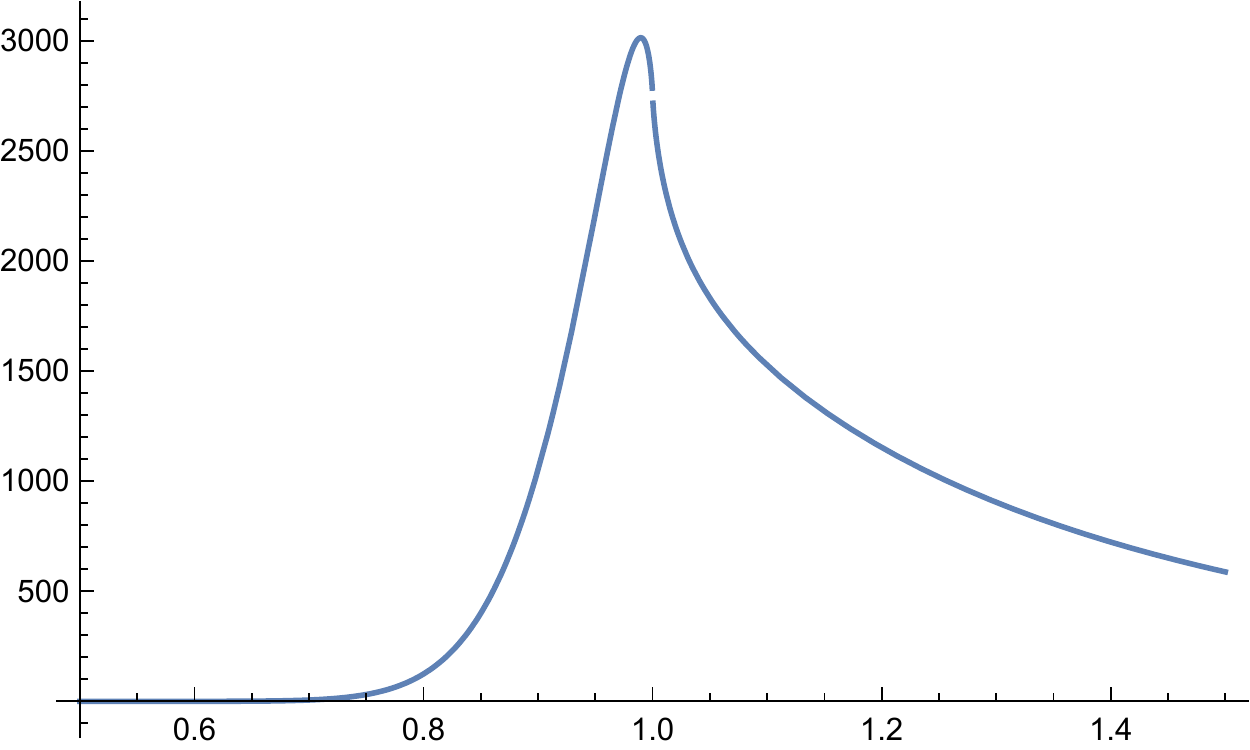}
\end{figure}

The graph displays some interesting features. First, the derivative of $\Delta S$ is infinite at $T=T_c$. This is expected and simply reflects the second order deconfinement transition. Recall, that neglecting the vortex effects, i.e. setting $m_1=0$, shifts the transition temperature to half the value  \cite{dkkt}. Thus it is the nontrivial $m_1$ dependence that drives the transition at $T_c$. 

The second interesting observation is that $\Delta S$ is positive. Thus the presence of vortices adds disorder to the system. 
Perhaps this is not entirely surprising. Our terminology so far has been  slightly misleading. We have been  referring to the $m_1=0$ limit as absence of vortices, whereas in fact the vortices are always present in the theory. In the limit $m_1\rightarrow 0$
the vortices do not disappear from the ensemble but rather the monopole-instantons are disallowed. This means that in this limit the magnetic symmetry is enhanced from $Z_2$ to $U(1)$. In general more symmetry does indeed mean more order, or less entropy. Thus at $m_1\ne 0$ the entropy is increased.
Finally, at high temperature, as expected $\Delta S\rightarrow 0$. This decrease is however rather slow 
\be
\Delta S(m_1))\rightarrow _{T\rightarrow \infty} \frac{V M_W m_1^2}{2\pi T}\sim 1/T^3
\ee

Recall that in the Yang-Mills theory $\Delta S<0$, see \cite{karsch}. Thus in this respect the 2+1 Georgi-Glashow model is different from 3+1 QCD.

\subsection{Free energy and Entropy due to charged particles}
In the previous subsection we have calculated the free energy and entropy excess due to effects of magnetic monopoles. In the imaginary time formalism, the charged particles appear in the thermal ensemble in a quite similar way. It is therefore amusing to see what is their contribution to $F$ and $S$. Excluding charged particles from the ensemble obviously is equivalent to setting $m_2=0$.

This calculation is very similar to the one performed in the previous subsection.
Starting from free energy, we obtain the analogue of eq.(\refeq{fm1})

\begin{align}
\begin{split}
\frac{\del F}{\del m_2}&=-\frac{TV}{2}\frac{1}{(2\pi)^2}(2\pi)\int_0^\Lambda \frac{dp^2}{2}\Big\{\frac{2(m_1+m_2)}{p^2+(m_1+m_2)^2}+\frac{2(m_1-m_2)}{p^2+(m_1-m_2)^2} \Big\} \\
&=TV\frac{1}{2\pi}\Big\{ -(m_1-m_2)\ln |m_1-m_2|+(m_1+m_2) \ln |m_1+m_2|-2m_2\ln T  \Big\}
\end{split}
\end{align}

This gives the  additional free energy due to the presence of charged particles

\begin{align}\label{Fm2}
\Delta F(m_2)=\frac{TV}{4\pi}\Big\{ (m_1-m_2)^2\ln|m_1-m_2|+(m_1+m_2)^2\ln|m_1+m_2|-2m_1^2\ln(m_1)-2m_2^2\ln T-m_2^2  \Big\}
\end{align}

And an additional entropy (up to leading order in $T/M_W$):
\be\label{sm2}
\Delta S(m_2)=\frac{Vm_2^2}{2\pi T}\Big\{(k+1)\ln(1+k)+(1-k)\ln|1-k| +\ln \frac{m_1^2}{T^2}  \Big\}
\ee
This is shown in Fig.2.  This is quite different from the vortex contribution. In particular the entropy is dominated by the last term. It is negative and much larger than the rest of the terms in eq.(\ref{sm2}). In particular the fact that the derivative of $S$ diverges at $T=T_c$ is difficult to see on the scale of Fig.2. The negativity of the entropy may be surprising at first sight, since we are adding massive particles to the ensemble, which should increase entropy. However these particles have an effect of producing   more order in the distribution of vortices, since heavy charged bosons are in fact nonlocal solitons of the vortex field. This ordering effect must be what reduces the entropy once the heavy charged bosons are taken into account. 
\begin{figure}[h!]
	\caption{$\Delta S(m_2)/VM_We^{-8\pi M_W/g^2}$ as a function of $T/T_c$ at $M_W/T_c=10$.}
	\includegraphics[scale=0.8]{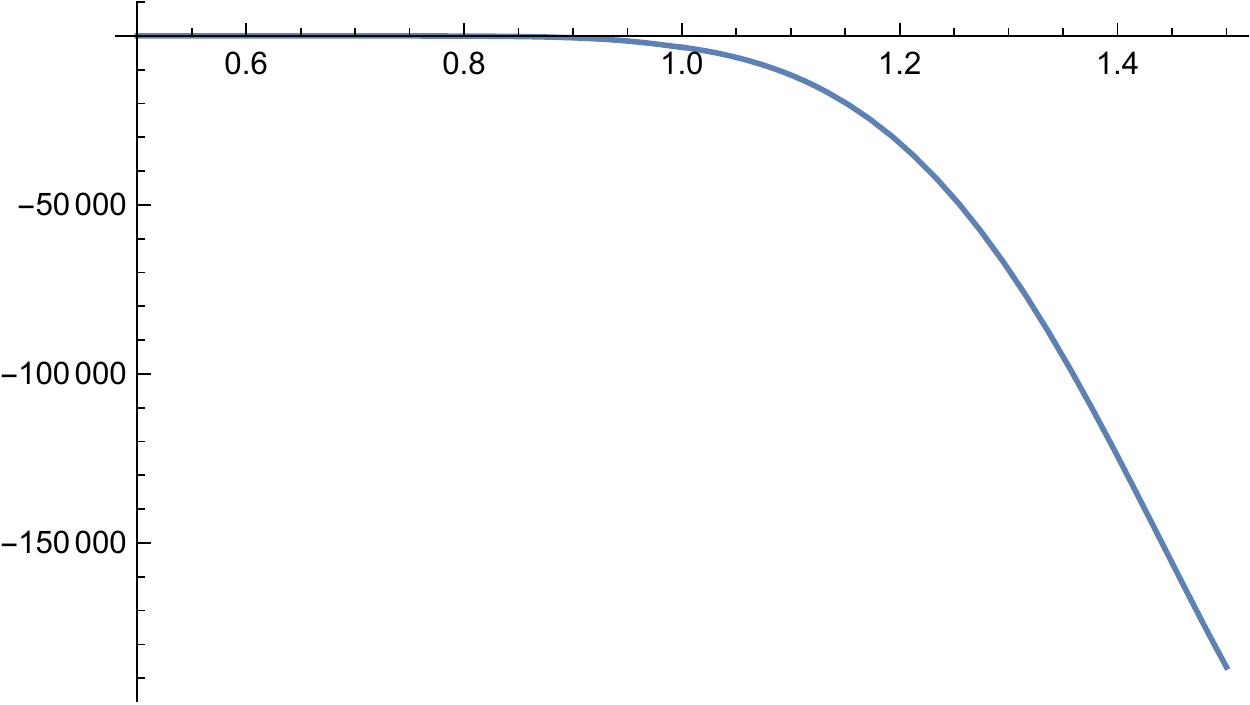}
\end{figure}

\subsection{The Free Energy}
Finally, using the previous expressions we can calculate the free energy of the theory. We use:
\begin{align}
\begin{split}
\Delta F(m_1)=F(m_1,m_2)-F(0,m_2)\\
\Delta F(m_2)=F(m_1,m_2)-F(m_1,0)
\end{split}
\end{align}
to find
\begin{align}
\begin{split}
\Delta F(m_1)-\Delta F(m_2)&=F(m_1,0)-F(0,m_2)\\
&=\frac{TV}{4\pi}\Big\{2m_1^2\ln m_1-2m_2^2\ln m_2+2(m_2^2-m_1^2)\ln T+(m_2^2-m_1^2)\Big\}
\end{split}
\end{align}

And we obtain

\begin{align}
F(m_1,m_2)=\frac{TV}{4\pi}\Big\{(m_1-m_2)^2\ln|m_1-m_2|+(m_1+m_2)^2\ln|m_1+m_2|-2(m_1^2+m^2_2)\ln T-(m_1^2+m_2^2)\Big\}+F(0,0)
\end{align}
The ``constant" $F(0,0)$ is nothing but the free energy of a free massless field, since this is the sole content of our model at vanishing $m_1$ and $m_2$. Finally we find

\begin{align}
F(m_1,m_2)=\frac{TV}{4\pi}\Big\{(m_1-m_2)^2ln\Big|\frac{(m_1-m_2)}{\Lambda}\Big|+(m_1+m_2)^2ln\Big|\frac{(m_1+m_2)}{\Lambda}\Big|+(cT^2-m_1^2-m_2^2)\Big\}
\end{align}

where c=3$\zeta(3)/2\pi$.

%\begin{align}
%F(m_1,m_2)=\frac{TV}{4\pi}\Big\{(m_1-m_2)^2\ln|m_1-m_2|+(m_1+m_2)^2\ln(m_1+m_2)+2(2T^2-m_1^2-m_2^2)\ln T+(2T^2-m_1^2-m_2^2)\Big\}
%\end{align}

\section{Conclusions.}
In this paper we have studied the 2+1 dimensional Georgi-Glashow model close to critical temperature. We used the fact that the generating functional of the model close enough to criticality is equivalent to a theory of noninteracting massive fermions. 

We have shown by explicit calculation in the fermionized  framework that the expectation value of the vortex operator ('t Hooft loop) vanishes above critical temperature, but is non vanishing below $T_c$. Conversely, the Polyakov loop vanishes below $T_c$ and is nonzero above $T_c$. We have also calculated the infrared asymptotic behavior of the correlators $\langle V^*(x)V(y)\rangle$ and $\langle V(x)V(y)\rangle$. We found that the two correlators are very close and if fact their leading behavior is the same
$\langle V^*(x)V(y)\rangle \approx \langle V(x)V(y)\rangle$. This means that in the near transition region the anomalous breaking of the magnetic $U(1)$ symmetry is an order one effect. This is very different form the situation at high temperatures, where the ratio of the two correlators is a large number and the anomalous breaking is small. 

Finally we have calculated the free energy and entropy of the model. We have shown that the main non perturbative effect, namely non vanishing  mass term for the vortex field leads to decrease of free energy, but  increase of entropy. In this respect this model is different from 3+1 dimensional Yang-Mills theory, where lattice data shows that non perturbative effects decrease both the free energy and entropy.

This suggests that important physical mechanisms at play in the two theories are different. Indeed confinement in the Georgi-Glashow model is ``abelian", in the sense that the low energy theory is effectively abelian and confining strings are classical and thick (the string width is much larger than the inverse of the square root of the string tension).
The confining strings in QCD are quantitatively quite different. The width of the confining string is presumably given by the inverse mass of the lightest glueball, which is much smaller than the inverse square of the string tension\footnote{For nice discussion of this issue see\cite{yung}.}.

In 2+1 GG model, the objects ``dual" to the thick confining strings are monopoles which are practically pointlike on the scale determined by critical temperature.  Thus their configurational entropy is large. One can speculate that in QCD the situation is reversed. The relevant topological objects (whatever they are), being dual to thin strings may themselves be fluffy and large. Configurational entropy of such objects is then relatively small, and their ordering effect on random field fluctuations may be large, leading to the overall decrease of entropy.

\begin{acknowledgements}
This research was supported by the DOE grant DE-FG02-13ER41989.
\end{acknowledgements}

\section{Appendix.}
\subsubsection{Cartesian to Polar Coordinates:}
\be  x=r\sin{\theta}, \mbox{ } y=-r \cos{\theta}. \ee

\subsubsection{Partial Derivatives:}
\begin{align}\begin{split}
 \partial_r = \sin{\theta} \partial_x - \cos{\theta}\partial_y,\ 	\	\	\ &\partial_\theta = r\cos{\theta}\partial_x + r \sin{\theta}\partial_y,\\
   \partial_x = \sin{\theta} \partial_r +\frac{1}{r} \cos{\theta}\partial_\theta,\	\	\	\  &\partial_y = -\cos{\theta}\partial_r +\frac{1}{r} \sin{\theta}\partial_\theta,\\
   \mbox{and}&\\
   \partial_x \pm i\partial_y=e^{\pm i\theta}&\left(\mp  i \partial_r + \frac{1}{r}\partial_\theta\right).
   \end{split}\end{align}

\subsubsection{Vector Potential:}
\be A_x \pm i A_y =  \frac{1}{r}e^{\pm i\theta} \label{aae}. \ee

\subsubsection{Covariant Derivative:}
\be D_x \pm i D_y = e^{\pm i\theta}\left(\mp i\partial_r +\frac{1}{r}\partial_\theta - \frac{i}{2r}\right) \label{dae}.\ee


\begin{thebibliography}{1}
\bibitem{svet} B. Svetitsky and L. Yaffe,  Nucl.Phys. B210 (1982) 423-447; B. Svetitsky, Phys.Rept. 132 (1986) 1-53 
\bibitem{karsch}  G. Boyd, J. Engels, F. Karsch, E. Laermann, C. Legeland, M. Lutgemeier, B. Petersson; Nucl.Phys. B469 (1996) 419-444 ; e-Print: hep-lat/9602007 
\bibitem{polyakov} A. Polyakov, Nucl.Phys. B120 (1977) 429-458
\bibitem{dkkt}  G.Dunne, I.Kogan, A. Kovner and B. Tekin, JHEP0101, 032 (2001) hep-th/0010201.
\bibitem{kr} A. Kovner and B. Rosenstein,Int.J.Mod.Phys. A7 (1992) 7419-7514; A.Kovner, Published in  *Shifman, M. (ed.): At the frontier of particle physics, vol. 3* 1777-1825 ; 
e-Print: hep-ph/0009138 
\bibitem{sonkovchegov} D.T. Son and Y. Kovchegov, JHEP 0301 (2003) 050 hep-th/0212230
\bibitem{boson1}S. Coleman, Phys. Rev. D11 (1975) 2088.
\bibitem{boson2}S. Mandelstam, Phys. Rev. D11 (1975) 3026.
\bibitem{chris} A. Smilga, Ann. Phys. 234, 1 (1994); Acta Phys. Polon., B25, 73 (1994);
\bibitem{chiral} See for example R. D. Pisarski and V. V. Skokov  Phys.Rev. D94 (2016) no.3, 034015; e-Print: arXiv:1604.00022 [hep-ph] 
\bibitem{jackiw}  R. Jackiw, P. Rossi, Nucl.Phys. B190 (1981) 681-691	
\bibitem{rob} A. Dumitru, Y. Hatta, J. Lenaghan,  K. Orginos and R. D. Pisarski;  Phys.Rev. D70 (2004) 034511, e-Print: hep-th/0311223 
\bibitem{shuryak} E. Shuryak and T. Schaeffer, Rev.Mod.Phys. 70 (1998) 323-426 ; e-Print: hep-ph/9610451
\bibitem{hight} I.Kogan, A. Kovner and B. Tekin, JHEP 0103 (2001) 021, e-Print: hep-th/0101171 
\bibitem{dk} G. Dunne and A. Kovner,  Phys.Rev. D82 (2010) 065014; e-Print: arXiv:1004.1075 [hep-ph]
\bibitem{yung} A. Yung, In *Shifman, M. (ed.): At the frontier of particle physics, vol. 3* 1827-1857

\end{thebibliography}
\end{document}